\documentclass[fleqn,usenatbib]{mnras}

\usepackage[T1]{fontenc}

\DeclareRobustCommand{\VAN}[3]{#2}
\let\VANthebibliography\thebibliography
\def\thebibliography{\DeclareRobustCommand{\VAN}[3]{##3}\VANthebibliography}

\usepackage{graphicx}	
\usepackage{amsmath}	
\usepackage{mathtools}
\usepackage{xcolor}
\usepackage[normalem]{ulem}

\usepackage{caption}
\usepackage{subcaption}
\usepackage{hyperref}

\usepackage{newtxtext,newtxmath}

\title[MHD waves in rotating photospheric flux tubes]{The effect of linear background rotational flows on magnetoacoustic modes of a photospheric magnetic flux tube}

\author[S. J. Skirvin et al.]{
S. J. Skirvin$^{1,2}$\thanks{E-mail: samuel.skirvin@kuleuven.be},
V. Fedun$^{2}$,
S. S. A. Silva$^{2}$, T. Van Doorsselaere$^{1}$, N. Claes$^{1}$, M. Goossens$^{1}$ and G. Verth$^{3}$
\\
$^{1}$Centre for mathematical Plasma Astrophysics, Mathematics Department, KU Leuven, Celestijnenlaan 200B bus 2400, B-3001 Leuven, Belgium. \\
$^{2}$Plasma Dynamics Group, Department of Automatic Control \& Systems Engineering, The University of Sheffield, Sheffield, S3 7RH, UK\\
$^{3}$Plasma Dynamics Group, Department of Mathematics and Statistics, The University of Sheffield, Sheffield, S3 7RH, UK\\
}

\date{Accepted XXX. Received YYY; in original form ZZZ}

\pubyear{2022}

\begin{document}
\label{firstpage}
\pagerange{\pageref{firstpage}--\pageref{lastpage}}
\maketitle

\begin{abstract}
Magnetoacoustic waves in solar magnetic flux tubes may be affected by the presence of background rotational flows. Here, we investigate the behaviour of $m=0$ and $m=\pm 1$ modes of a magnetic flux tube in the presence of linear background rotational flows embedded in a photospheric environment. We show that the inclusion of a background rotational flow is found to have little effect on the obtained eigensolutions for the axisymmetric $m=0$ sausage mode. However, solutions for the kink mode are dependent on the location of the flow resonance modified by the slow frequency. A background rotational flow causes the modified flow resonances to possess faster phase speeds in the thin-tube (TT) limit for the case $m=1$. This results in solutions for the slow body and slow surface kink modes to follow this trajectory, changing their dispersive behaviour. For a photospheric flux tube in the TT limit, we show that it becomes difficult to distinguish between the slow surface and fast surface kink ($m=1$) modes upon comparison of their eigenfunctions. 2D velocity field plots demonstrate how these waves, in the presence of background rotational flows, may appear in observational data. For slow body kink modes, a swirling pattern can be seen in the total pressure perturbation. Furthermore, the tube boundary undergoes a helical motion from the breaking of azimuthal symmetry, where the $m=1$ and $m=-1$ modes become out of phase, suggesting the resulting kink wave is circularly polarised. These results may have implications for seismology of magnetohydrodynamic waves in solar magnetic vortices.

\end{abstract}

\begin{keywords}
magnetohydrodynamics (MHD) -- waves
\end{keywords}



\section{Introduction}\label{Introduction}

High resolution observations have revealed, over the last few decades, that magnetohydrodynamic (MHD) waves are ubiquitous throughout the Sun's atmosphere \citep{nak1999, Asc1999, dep2007, Tomczyk_et_al_2007, Morton2015, grant2015, keys2018, stang2022, Bate2022}. Understanding MHD wave properties from a theoretical point of view is of the utmost importance in solar physics as these waves could contribute to the heating of local plasma and may also be used as a proxy to determine sub-resolution plasma properties. Additionally, observed MHD wave behaviour can be used to provide an estimate of the properties of local plasma that cannot be measured directly, for example the magnetic field in the corona. Furthermore, MHD waves may also be responsible in the formation of some observed phenomena, e.g. jets, in the solar atmosphere \citep{dep2004,RouppevanderVoort2007,scu2011}. 

The uniform cylindrical waveguide model \citep[see e.g.][]{wilson1979, Spr1979, edrob1983} has provided a foundation for analytical studies into MHD wave investigations under solar atmospheric conditions. These studies have shown that the trapped wave modes of a magnetic flux tube can be described by the number of nodes present in each geometrical direction. The axisymmetric sausage mode has zero nodes in the azimuthal direction $m=0$, whereas the (linearly polarised) non-axisymmetric kink mode has a single azimuthal node $m=1$ or $m=-1$. Whilst this uniform model is simplistic in nature, due to the uniform plasma considered inside and outside the waveguide, it has been modified over recent decades to model specific configurations which better match observational data \citep[see e.g.][]{TVD2004_curvature,verth2007,ErdFed2010,Anwar2021,rud2022}. The cylinder model can be further extended to incorporate additional physical environments that may be common configurations in magnetic flux tubes, such as including a background rotational flow. Magnetic flux tubes with background rotational flows are a common configuration observed in, e.g., flux tubes rooted in intergranular lanes, solar tornadoes and spicules \citep{Bonet2010, Wedemeyer2012, Tzi2018, Shetye2019}. Such structures display dynamic characteristics and excite a wide range of MHD waves which couple different layers of the solar atmosphere. As a result, these structures act as a natural conduit for the transfer of mass, momentum and energy throughout the solar atmosphere. Furthermore, rotational flows naturally appear in numerical MHD simulations of regions in the solar atmosphere with vortex drivers \citep[see, e.g.][]{fed2011, fedshe2011, shel2011, shelyag2012, Shel2013, Gonzalez-Aviles_et_al_2017, gon2018, sno2018} and also in magnetoconvection simulations \citep{Yadav2020, Yadav2021, Silva2021}. 

The presence of a background rotational flow would manifest itself as a non-zero azimuthal component of the background velocity field vector. Considering $m=\pm 1$ transverse kink modes, the effect of introducing a background rotational flow into the model would break the symmetry of the system with respect to the direction of azimuthal wave propagation. Similarly, it has been shown that the inclusion of a steady vertical background plasma flow, aligned with the magnetic field, can obtain solutions which also break the symmetry of the forward and backward propagating waves \citep{nak1995,terra2003, Soler2009, Skirvin2022}. The presence of the longitudinal flow results in a Doppler-shifted wave frequency, depending upon the amplitude of the flow and the vertical wavenumber, which may shift certain wave modes into different physical regimes. Furthermore, a background longitudinal flow may also affect the wavelength and damping length of resonant absorption for kink modes, causing more efficient damping of backward propagating modes compared to forward propagating modes \citep{Soler2011}.

Circularly polarised kink modes, previously observed in chromospheric magnetic elements \citep{stang2017} and sunspots \citep{Jess2017}, have also been studied recently by \citet{magyar2022} where the authors discuss the differences that resonant absorption and phase mixing have on linearly and circularly polarised kink waves. \citet{magyar2022} also conclude, upon analysis of the Doppler signatures for both polarisation states, that there is very little difference between the two polarisation's in Doppler observations. Circularly polarised kink waves have been shown to develop in coronal loops with twisted magnetic fields \citep{ter2012,rud2015}. 

Previous analytical studies have investigated the stability status of rotating flux tubes, as an azimuthal velocity shear across the waveguide boundary may be susceptible to the Kelvin-Helmholtz instability (KHI) \citep{sol2010, zaq2015, zhe2019}. However, these studies assume zero plasma-$\beta$ and focus on coronal conditions only, ignoring the slow magnetoacoustic modes entirely. The stability of a magnetic flux tube with a linear background magnetic twist and rotational flow component was studied by \citet{Cher2018} who found that the $m=0$ sausage mode becomes unstable for azimuthal flow speeds that create a centrifugal force which can overcome the magnetic tension, whereas, the $m=1$ kink mode can only become unstable for sufficiently large values of longitudinal (axial) flow speed. Whilst the stability studies mentioned above investigate the susceptibility of the KHI to various wave modes in magnetic flux tubes with background flows, we stress that it is not the goal of this study. The primary aim of the present work, however, is to provide a description of how $m=0$ and $m=\pm 1$ wave modes would manifest themselves in solar magnetic flux tubes, under photospheric conditions, in the presence of background rotational flows and to aid mode interpretations of observational and numerical data.

This paper is presented as follows, in Section \ref{Method} the governing equations describing a magnetic flux tube in the presence of a background rotational flow is presented, along with a brief description of the numerical eigensolver implemented in this work. Section \ref{twisted_flow} presents the results of an investigation into the effect of a linear profile of background rotational flow on the properties of magnetoacoutsic waves under photospheric conditions. Section \ref{twisted_flow} looks closely at the obtained eigenvalues, one dimensional eigenfunctions and two/three dimensional visualisations of the kink mode in the presence of a background rotational flow. Finally, in Section \ref{conclusions}, we summarise the findings presented in this work and discuss avenues of future research.

\section{Method}\label{Method}
The ideal MHD equations adopted in this study are:
 \begin{align}
    \frac{d\rho}{dt}+\rho\nabla\cdot\textbf{v}&=0, \label{continuity} \\
    \rho\left(\frac{d\textbf{v}}{dt}\right)&=-\nabla p+\frac{1}{\mu_0}\left(\nabla\times\textbf{B}\right)\times\textbf{B}, \label{momentum} \\
    \frac{d}{dt}\left(\frac{p}{\rho^\gamma}\right)&=0, \label{energy} \\ 
    \frac{\partial\textbf{B}}{\partial t}&=\nabla\times\left(\textbf{v}\times\textbf{B}\right), \label{induction} \\
    \nabla \cdot \textbf{B} &= 0, \label{divB}
 \end{align}

where $\rho$, $\textbf{v}$, $p$, $\textbf{B}$, $\gamma$ and $\mu$ denote plasma density, plasma velocity, plasma pressure, magnetic field, ratio of specific heats (taken $\gamma = 5/3$) and the magnetic permeability respectively. We consider a cylindrical geometry $(r, \varphi, z)$ where the initial equilibrium has a radially spatially dependent form of the velocity field vector $\textbf{v}_0 = (0, v_{\varphi}(r), 0)$. In our model the magnetic field is taken to be straight and uniform with no background axial plasma flow. In the lower solar atmosphere it should be noted that magnetic flux tubes possess significant non-vertical magnetic field as a result of flux tube expansion to maintain pressure balance, however, this effect can be considered to be negligible in the current study due to the analysis in the local plasma environment. Since the equilibrium quantities depend on $r$ only, the perturbed quantities can be Fourier-analysed with respect to the ignorable coordinates $\varphi, z$ and time $t$ and put proportional to: 
\begin{equation*}
\text{exp}\left[i\left(m\varphi + kz - \omega t\right)\right],
\end{equation*}
where $m$ is the azimuthal wave number, $k$ the vertical wavenumber and $\omega$ the wave frequency.

After linearising Equations (\ref{continuity})-(\ref{divB}), we arrive at a system of two differential equations containing the total pressure perturbation $\hat{P}_T$ and the radial displacement perturbation $r\hat{\xi}_r$ \cite[see e.g.][]{sak1991, goo1992}, which can be written as:

\begin{align}
    D\frac{d}{dr}\left(r\hat{\xi}_r\right)&=C_1 r\hat{\xi}_r - C_2 r \hat{P}_T, \label{rxi_r_diff} \\
    D\frac{d\hat{P}_T}{dr}&=C_3 \hat{\xi}_r - C_1 \hat{P}_T, \label{P_diff}
\end{align}
where,
\begin{align}
    D &= \rho\left(c^2+v_A^2 \right)\left(\Omega^2-k^2v_A^2\right)\left(\Omega^2-k^2c_T^2\right), \label{D} \\
    \Omega &= \omega - \frac{m}{r}v_{\varphi}, \label{Omega} \\
    c^2 &= \frac{\gamma p}{\rho},\ \ \ \ v_A^2 = \frac{B_z^2}{\mu\rho}, \ \ \ \ c_T^2 = \frac{v_A^2 c^2}{\left(c^2 + v_A^2\right)}, \\
    C_1 &= Q\Omega^2 - 2m\left(c^2+v_A^2\right)\left(\Omega^2-k^2c_T^2\right)\frac{T^2}{r^2}, \label{C_1} \\
    C_2 &= \Omega^4 - \left(c^2+v_A^2\right)\left(\frac{m^2}{r^2}+k^2\right)\left(\Omega^2-k^2c_T^2\right), \label{C_2} \\
    C_3 &= D\left\{\rho\left(\Omega^2-k^2v_A^2 \right) + r\frac{d}{dr}\left[- \rho\left(\frac{v_{\varphi}}{r}\right)^2\right]\right\} + \\ \nonumber &+ Q^2 - 4\left(c^2+v_A^2 \right)\left(\Omega^2-k^2c_T^2\right)\frac{T^2}{r^2}, \\
    Q &= -\left(\Omega^2-k^2v_A^2\right)\frac{\rho v_{\varphi}^2}{r}, \label{Q} \\
    T &= \rho \Omega v_{\varphi}. \label{T}
\end{align}
It should be noted here that we assume $B_{\varphi} = v_z = 0$,  which simplifies the fully inclusive ($B_{\varphi} = B_{\varphi}(r), B_z=B_z(r),  v_z = v_z(r), \rho = \rho(r)$) set of equations previously noted in literature by \citet{goo1992}. Quantities $c^2$, $v_A^2$, and $c_T^2$ define the squares of the local sound speed, Alfv\'{e}n speed and cusp (tube) speed respectively. The quantity $\Omega$ represents the Doppler shifted frequency as a result of the background rotational plasma flow.

Equations (\ref{rxi_r_diff})-(\ref{P_diff}) can be combined to create a single differential equation in either $r\hat{\xi}_r$:
\begin{equation}\label{rxi_r_diff_eqn}
    \frac{d}{dr}\left[ f(r)\frac{d}{dr}\left(r\hat{\xi}_r\right)\right] - g(r)\left(r\hat{\xi}_r\right) = 0,
\end{equation}
where,
\begin{equation}
f(r) = \frac{D}{rC_2},
\end{equation}
\begin{equation}
g(r) = \frac{d}{dr}\left(\frac{C_1}{rC_2}\right) - \frac{1}{rD}\left(C_3 - \frac{C_1^2}{C_2}\right),
\end{equation}
or $\hat{P}_T$:
\begin{equation}\label{P_diff_eqn}
    \frac{d}{dr}\left[ \Tilde{f}(r)\frac{d\hat{P}_T}{dr}\right] - \Tilde{g}(r)\hat{P}_T = 0,
\end{equation}
where,
\begin{equation}
\Tilde{f}(r) = \frac{rD}{C_3},
\end{equation}
\begin{equation}
\Tilde{g}(r) = -\frac{d}{dr}\left(\frac{rC_1}{C_3}\right) - \frac{r}{D}\left(C_2 - \frac{C_1^2}{C_3}\right).
\end{equation}
For a non-uniform plasma, the governing Equations (\ref{rxi_r_diff})-(\ref{P_diff}) possess regular singularities where the wave frequency matches the local characteristic frequencies at:
\begin{equation}\label{alfven_flow_continuum}
    \omega = \frac{m}{r}v_{\varphi}(r) \pm k v_A,
\end{equation}
\begin{equation}\label{cusp_flow_continuum}
    \omega = \frac{m}{r}v_{\varphi}(r) \pm k c_T.
\end{equation}
Equations (\ref{alfven_flow_continuum}) and (\ref{cusp_flow_continuum}) define the flow continua modified by the local Alfv\'{e}n ($kv_A$) and slow ($kc_T$) frequencies, respectively. In ideal MHD, the wave solutions existing inside the continua, with positions given by Equations (\ref{alfven_flow_continuum}) and (\ref{cusp_flow_continuum}), are known as `quasi-modes' where the wave frequency becomes a complex quantity \citep{DeGroof2000, goe2004, Geer2022}. The nature of the solutions lying inside the continua is not discussed in the present study and instead will be the focus of future work.

Both Equations (\ref{rxi_r_diff_eqn}) and (\ref{P_diff_eqn}) have no known closed form analytical solutions, without making assumptions that somehow reduce the mathematical complexity. Therefore, investigating the properties of wave modes propagating within an equilibrium which is non-uniform must be done numerically. The numerical approach used in this study is based on the eigensolver applied in \citet{Skirvin2021, Skirvin2022} for non-uniform magnetic slabs and non-uniform flux tubes, respectively. The eigensolver implements the numerical shooting and bisection methods whilst also relying on fundamental properties of the sausage and kink modes. The numerical shooting method solves Equations (\ref{rxi_r_diff_eqn}) and (\ref{P_diff_eqn}) ensuring continuity of $\hat{P}_T$ and $\hat{\xi}_r$ across the boundary of the flux tube. This technique has been applied before in solar physics by, e.g. \citet{Tirry1996, Pinter1998, Andries2000, tar2002, Tar2003}. These studies also utilise the jump conditions introduced by \citet{sak1991, goo1992} to deal with the regular singularities that appear in the governing equations, allowing the authors to study the resulting quasi-modes. However, following the primary objective of the present study, we only consider eigenmodes with real valued wave frequency and wavenumber.

\section{Magnetic flux tube in the presence of a linear background rotational flow}\label{twisted_flow}
In this section, a magnetic flux tube in the presence of a linear rotational background flow is investigated. For all cases the magnetic flux tube is otherwise uniform such that the equilibrium plasma density and magnetic field is constant across the flux tube. A profile comparable to the magnetic twist profile incorporated by \citet{ErdFed2007} and \citet{ErdFed2010} is chosen but applied to the azimuthal velocity field component $v_{\varphi}$ instead. A rotational flow can be either clockwise or counter-clockwise in the reference frame relative to the observer. The only difference between a clockwise rotational flow and an anti-clockwise rotational flow will be the sign in front of $v_{\varphi}$ and the direction of shifted wave frequency relative to the flow. In the following sections, a magnetic flux tube is presented with an equilibrium azimuthal flow component, acting in a counter-clockwise direction, which takes the form: 
\begin{equation}\label{flow_eqn}
v_{\varphi}(r) = A\left(\frac{r}{a}\right)^\alpha, 
\end{equation}
where $A$ is the amplitude of the rotational flow, $\alpha$ is the parameter (exponent) dictating the radial profile of the rotational flow and $a$ indicates the location of the boundary of the flux tube, taken to be $a=1$ in our study. The case when $\alpha=1$ for example corresponds to a linear rotational flow, which will be the focus of this study, (see e.g. Figure \ref{Twisted_flow_Equilibrium_profiles_linear_photospheric}). It can be shown that when $v_{\varphi}$ is linear, Equations (\ref{rxi_r_diff})-(\ref{T}) simplify, and in many cases the dependence on the radial coordinate is removed. The rotational flow is constant with height $z$ in all cases considered in this work. Obtaining an equilibrium in a magnetic cylinder with a background rotational velocity component is not as mathematically simple as the scenario of a uniform magnetic cylinder. In order to maintain total pressure balance across the waveguide the following expression must be satisfied \citep{goo2011}:
\begin{equation}\label{rot_flow_pressure_balance}
    \frac{d}{dr}\left( p + \frac{B_{0z}^2}{2\mu} \right) = \frac{\rho v_{\varphi}^2(r)}{r} = \rho A^2 r.
\end{equation}
Integration of Equation (\ref{rot_flow_pressure_balance}) yields:
\begin{equation}\label{integrate_pressure_balance}
    p + \frac{B_{0z}^2}{2\mu} = \rho A^2 \frac{r^2}{2},
\end{equation}
where the constant of integration is absorbed into the gas pressure term, $p$, and corresponds to the plasma pressure on the axis of the cylinder where the amplitude of the flow is zero \citep[see e.g.][]{Cher2018}. Under the photospheric conditions considered in this work, the total pressure balance is achieved by an increase in temperature to balance the increase in azimuthal flow amplitude towards the boundary of the flux tube. For configurations where the amplitude of the rotational flow is weak (e.g. $A < 0.5c_i$), then the change in spatial behaviour of the plasma pressure and temperature is small, but must not be dismissed. 

The presence of a background rotational flow not only modifies the equilibrium pressure balance relationship, but also affects the continuity conditions on the boundary of the waveguide. Considering a magnetic flux tube in the presence of a background rotational flow, the resulting boundary continuity conditions state:
\begin{align}
    \label{rotflow_boundary_condition_xi} \hat{\xi}_{re}\Bigr\rvert_{r = a} &= \hat{\xi}_{ri}\Bigr\rvert_{r = a}, \\
    \label{rotflow_boundary_condition_PT} \hat{P}_{Te}\Bigr\rvert_{r = a} &= \left(\hat{P}_{Ti} + \frac{\rho_{0i} v_{\varphi}^2}{a}\hat{\xi}_{ri} \right)\Bigr\rvert_{r = a}.
\end{align}
The change in boundary conditions are accounted for in the numerical eigensolver, and a pair of eigenvalues will only be retrieved for values satisfying the above conditions for each respective case study.

Finally, when the background rotational flow is taken to be linear with respect to the radial direction, Equations (\ref{alfven_flow_continuum}) and (\ref{cusp_flow_continuum}) become:

\begin{align}
    \label{linear_alfven_flow_continuum} \omega &= Am \pm k v_A, \\
    \label{linear_cusp_flow_continuum} \omega &= Am \pm k c_T.
\end{align}
Equations (\ref{linear_alfven_flow_continuum}) and (\ref{linear_cusp_flow_continuum}) no longer define continuum regions as the resonant positions no longer cover a range of frequencies, rather they define singular resonant locations which correspond to the flow resonance position modified by the Alfv\'{e}n and slow frequencies respectively. The locations of these resonant positions depend heavily on the amplitude of the rotational flow in the setup presented in this work.

\subsection{Linear rotating magnetic flux tube under photospheric conditions}

\begin{figure}
    \centering
    \includegraphics[width=7.6cm]{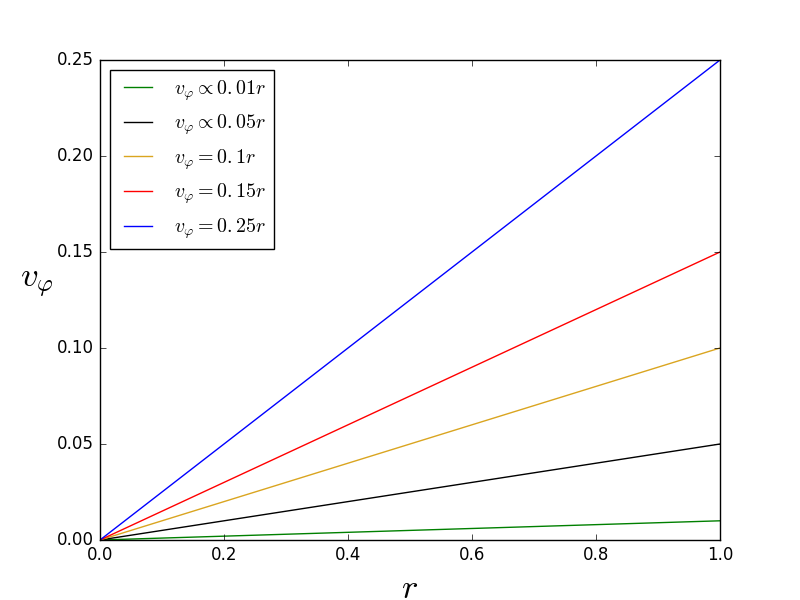}
    \caption{Equilibrium background rotational flow profiles for cases with increasing amplitude for a photospheric cylinder. In all cases the profiles are linear with respect to spatial coordinate $r$ (i.e. $\alpha=1$). The amplitude of the rotational flow increases linearly up to a boundary value of $A=0.01$ (green line), $A=0.05$ (black line), $A=0.1$ (yellow line), $A=0.15$ (red line) and $A=0.25$ (blue line). The boundary of the flux tube is located at $r=1$.}
    \label{Twisted_flow_Equilibrium_profiles_linear_photospheric}
\end{figure}

In this section, a magnetic flux tube under photospheric conditions ($v_{Ae} < c_{i} < c_e < v_{Ai}$) in the presence of a linear background rotational flow is investigated. For all photospheric cases in this work, the numerical values correspond to $c_i=1$, $c_e=1.5c_i$, $v_{Ai}=2c_i$, $v_{Ae}=0.5c_i$ and $\rho_i=1$. This choice of equilibrium parameters results in a density contrast between the internal and external plasma to be roughly $\rho_i/\rho_e = 0.567$. The background velocity vector inside the waveguide can be written as $\mathbf{v}_{0i} = (0, Ar, 0)$. The flow outside the cylinder is zero which results in a velocity shear across the cylinder boundary at $r=a$, however the value of $A$ is chosen to be small and both sub-Alfv\'{e}nic, sub-sonic and below the threshold for the Kelvin-Helmholtz instability \citep{sol2010}. This choice of amplitude $A$ also agrees with observed values of photospheric flows when compared with the local sound/Alfv\'{e}n speed \citep{Bonet2008}. Shown in Figure \ref{Twisted_flow_Equilibrium_profiles_linear_photospheric} are the linear profiles of background rotational flow considered in this section. In all cases the flow amplitude is proportional to the radial distance from the center of the flux tube, at $r=0$, up to the boundary at $r=1$, however the amplitude is allowed to vary.

\subsubsection{Eigenvalues}
   \begin{figure*}
    \centering
    \begin{subfigure}{.49\textwidth}
        \centering
        \includegraphics[width=8.cm]{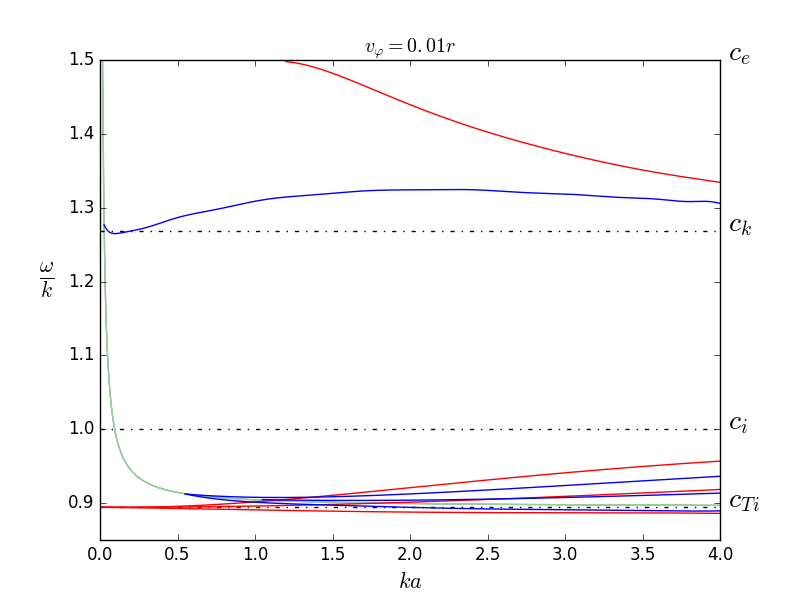}
        \caption{}
        \label{linear_flow_v001_dd}
   \end{subfigure}
   \begin{subfigure}{.49\textwidth}
        \centering
        \includegraphics[width=8.cm]{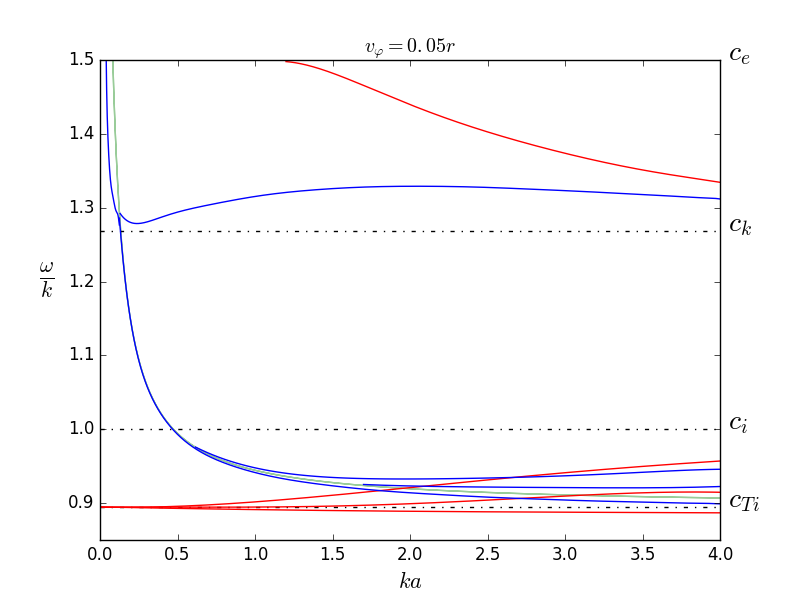}
        \caption{}
        \label{linear_flow_v005_dd}
    \end{subfigure}
      \begin{subfigure}{.49\textwidth}
        \centering
        \includegraphics[width=8.cm]{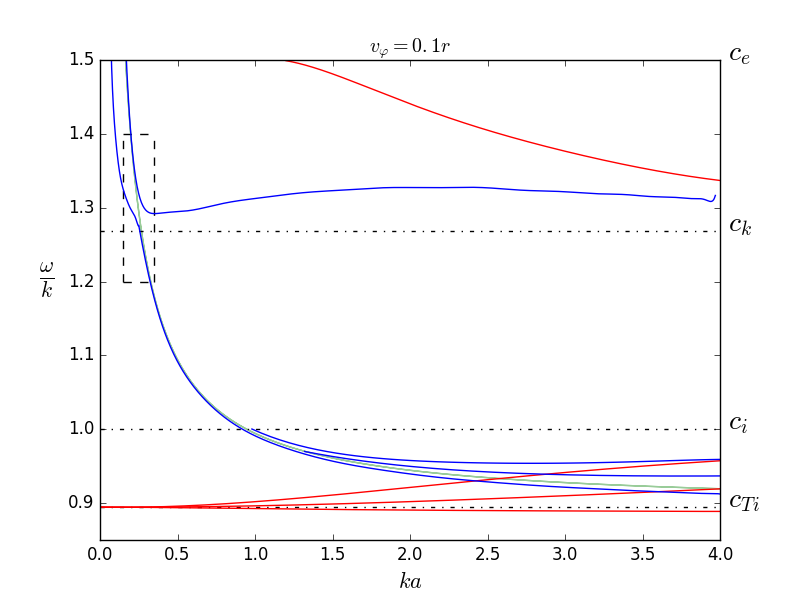}
        \caption{}
        \label{linear_flow_v01_dd}
    \end{subfigure} 
   \begin{subfigure}{.49\textwidth}
        \centering
        \includegraphics[width=8.cm]{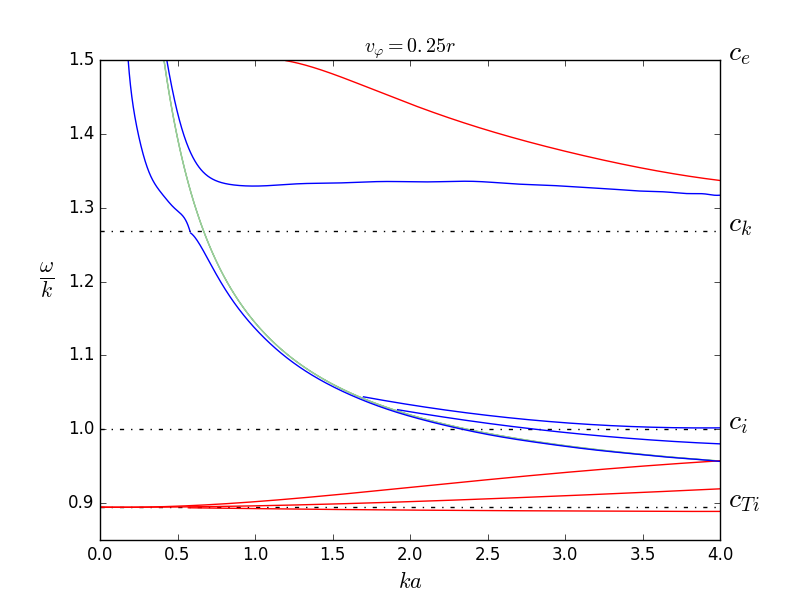}
        \caption{}
        \label{linear_flow_v025_dd}
    \end{subfigure} 
    \caption{Dispersion diagrams for a photospheric cylinder with a linear background rotational flow of varying flow amplitudes. The different cases with varying amplitude, displayed on the top of each panel, are shown corresponding to those in Figure \ref{Twisted_flow_Equilibrium_profiles_linear_photospheric}. The red curves indicate solutions for the $m=0$ sausage mode and the blue curves show the $m=+1$ kink mode solutions. The green curve in all cases shows the flow resonance locations modified by the slow frequency as given by Equation (\ref{linear_cusp_flow_continuum}). Dash-dotted lines indicate the equilibrium characteristic kink speed, $c_k$ sound speed, $c$, and tube speed, $c_{T}$, internal (subscript `i') and external (subscript `e') to the flux tube. The dashed box region shown in Figure \ref{linear_flow_v01_dd} is investigated in more detail in Figure \ref{linear_vrot_eigenfunction_comparison}.}
    \label{linear_flow_disp_diags_all_sols}
   \end{figure*}
Figure \ref{linear_flow_disp_diags_all_sols} highlights the change in eigenvalues for the different cases of flow profiles considered in Figure \ref{Twisted_flow_Equilibrium_profiles_linear_photospheric}. For now, we focus on the forward propagating wave modes ($\omega/k > 0$). The axisymmetric $m=0$ sausage mode appears unaffected, or at least not significantly affected, by the presence of the background flow, which is a result of the $m=0$ mode being the axisymmetric mode in the azimuthal direction. As a result of this, the rotational flow does not break the symmetry of this mode. This can also be seen analytically, by setting $m=0$ in Equations (\ref{rxi_r_diff})-(\ref{T}), the majority of terms containing the presence of the (linear) background flow are removed when $m=0$. Furthermore, when $m=0$, the resonant locations described by Equations (\ref{linear_alfven_flow_continuum}) and (\ref{linear_cusp_flow_continuum}) reduce to the resonant positions corresponding to a uniform magnetic flux tube \citep{edrob1983}. Whilst Equations (\ref{rxi_r_diff})-(\ref{T}) do contain some terms, for example in the variables $Q$ and $T$, which are dependent on the background rotational flow and that remain for the $m=0$ solution, we do not observe any significant modification to the obtained eigenvalues for the sausage mode in Figure \ref{linear_flow_disp_diags_all_sols} when compared to the analytical solutions for the uniform magnetic flux tube, although very minor effects can be seen when the amplitude of the rotational flow is sufficiently large ($A>0.25$). Investigating this result both analytically and numerically may be a focus of future work.

Conversely, there is a considerable effect on the $m=1$ kink mode solutions due to the presence of a background rotational flow. In the long wavelength (thin-tube) limit, the phase speeds of the slow body and slow surface kink modes tend to an infinite phase speed, similar to the case study of a linear background magnetic twist \citep{ErdFed2010}, where they may even enter the leaky regime. As the amplitude of the azimuthal flow is increased, the corresponding phase speeds of the slow kink modes also increases for all values of wavenumber. The flow resonance locations modified by the slow frequency given by Equation (\ref{linear_cusp_flow_continuum}) is shown by the green line in Figure \ref{linear_flow_disp_diags_all_sols}. The slow body and slow surface kink modes follow this resonance curve in the long wavelength limit and even undergo an avoided crossing \citep{abd1990, mather_2016, all2017} where the slow surface mode approaches the fast surface mode. This avoided crossing implies a transfer of properties between the fast and slow surface modes. In the long wavelength limit it may not be appropriate to refer to the slow surface mode as a slow mode anymore as, in the reference frame of the observer, it possesses phase speeds similar to that of the fast surface mode. Therefore, using the observed phase speeds alone may introduce some difficulty when distinguishing between the fast and slow surface kink modes in rotating photospheric flux tubes. It may be more appropriate to differentiate between the two modes by comparing their eigenfunctions both parallel and perpendicular to the magnetic field, which is discussed in more detail in Section \ref{Subsection_eigenfunctions}. In the case of a uniform photospheric cylinder, the slow body modes tend toward $c_{Ti}$ in the long wavelength limit \citep{edrob1983, Priest2014}, however with the inclusion of a background rotational flow, these modes now encounter the modified flow resonance point where they may become resonantly damped. Furthermore, the dispersive nature of the fast surface kink mode is also modified by the presence of a background rotational flow. In a uniform photospheric flux tube, the phase speed of the fast surface mode tends to the kink speed in the long wavelength limit \citep{edrob1983}. However, similar to the slow modes, the fast surface mode encounters the modified flow resonance in the long-wavelength limit, where the phase speed of this mode also appears to increase sharply depending on the amplitude of the flow. The fast mode may also enter the leaky regime in the long-wavelength limit under the assumed equilibrium configuration, which may be important in the context of solar observations.

\begin{figure*}
    \centering
    \includegraphics[width=15cm]{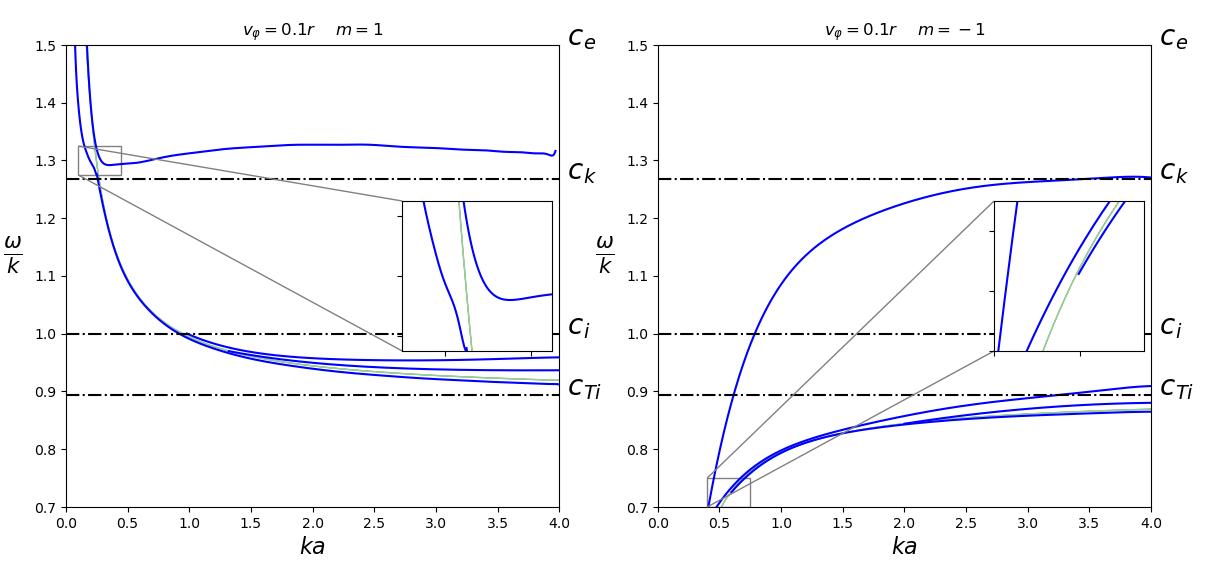}
    \caption{The dispersion diagrams showing the obtained solutions for the $m=1$ and $m=-1$ modes denoted by the blue curves. The green curve shows the behaviour of Equation (\ref{linear_cusp_flow_continuum}) for both cases of $m=\pm1$, respectively. The respective zoom in plots highlight the difference in the behaviour of the various category of modes for different values of $m$.}
    \label{m_comparison}
\end{figure*}

Of course, the kink mode possesses an azimuthal wavenumber which can be either positive or negative. Granular buffeting in the lower solar photosphere due to convective motions beneath the solar surface, may excite both $m=1$ and $m=-1$ kink modes. In the uniform cylinder model of a solar waveguide, the `traditional' kink mode is considered to display signatures that resemble a periodic transverse displacement of the waveguide. This is because, due to the symmetry of the model, the opposite rates of rotation set up a standing wave in the azimuthal direction. However, introducing a background rotational flow breaks this azimuthal symmetry and, as a result, the $m=1$ and $m=-1$ modes are expected to behave differently and the resulting observed mode will no longer be a standing wave in the azimuthal direction. In Figure \ref{m_comparison}, we show the obtained wave solutions for the $m=1$ and $m=-1$ modes in a photospheric flux tube with a linear background rotational flow of amplitude $A=0.1$. For the case when the kink wave is rotating with the flow, the phase speed of the fast surface kink mode increases in the long-wavelength limit and enters the leaky regime for small $ka$. This suggests that the fast kink surface mode, when propagating in a thin waveguide with the background rotational flow, should not be seen in observations of rotating photospheric structures, for example in magnetic bright points or flux bundles rooted within intergranular lanes. However, the obtained solutions for the fast surface mode are different for the two cases when $m=1$ and $m=-1$. This is due to the $m=1$ mode rotating in the same direction as the background flow, in our specific configuration. Therefore, the $m=1$ mode constructively interferes with the background rotational flow, increasing its phase velocity, whereas the $m=-1$ mode, rotating against the flow, destructively superimposes with the flow, hence lowering its phase velocity. The most notable difference between the case when $m=1$ and $m=-1$ can be seen by the dispersive behaviour of the slow surface and body modes. In both cases, the slow modes follow the trajectory of the flow resonance locations modified by the slow frequency as given by Equation (\ref{linear_cusp_flow_continuum}), however for the case when $m=-1$, this curve decreases with decreasing $ka$. As a result, the difference in phase speed for a given wavenumber between $m=1$ and $m=-1$ is greater for the slow modes, with a more dramatic difference seen as $ka$ approaches zero. This example of the phase speed difference between the forward propagating slow body and surface modes, for $m=1$ and $m=-1$, demonstrates how the background flow breaks the symmetry of the kink mode.

\subsubsection{Eigenfunctions}\label{Subsection_eigenfunctions}

\begin{figure}
    \centering
    \includegraphics[width=.49\textwidth]{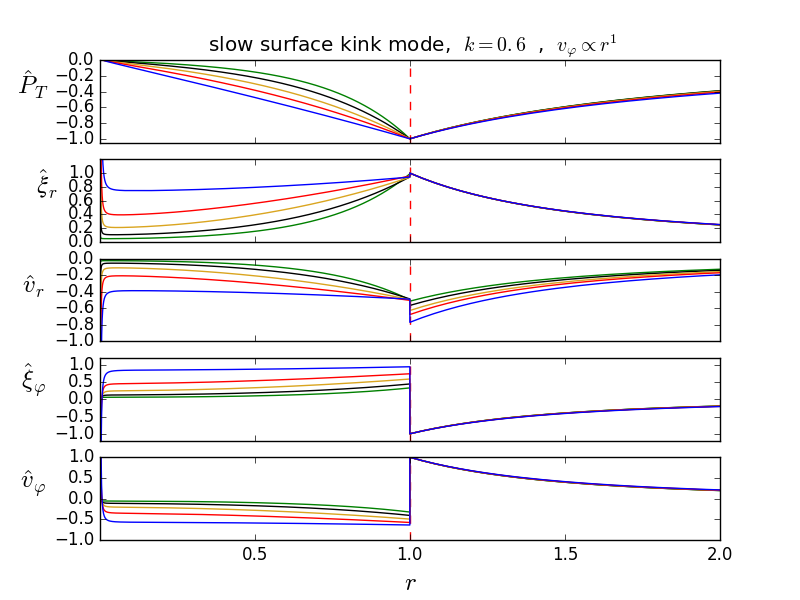}
    \caption{The resulting eigenfunctions for the slow surface kink mode ($m=1$) for all linear cases of rotational flow with the colour scheme consistent with Figure \ref{Twisted_flow_Equilibrium_profiles_linear_photospheric}. A wavenumber value of $k=0.6$ was chosen for all plots.}
    \label{Twisted_flow_kink_eigenfunctions_photospheric}
\end{figure}
Shown in Figure \ref{Twisted_flow_kink_eigenfunctions_photospheric} are the spatial eigenfunctions for the slow surface kink mode at a fixed wavenumber $ka=0.6$ for varying rotational flow amplitudes with $m=1$. The colour scheme shown in the eigenfunctions is consistent with that for the rotational flow profiles shown in Figure \ref{Twisted_flow_Equilibrium_profiles_linear_photospheric}. It can be seen that increasing the amplitude of the equilibrium linear rotational flow, changes the spatial behaviour of the observable eigenfunctions. For the case of $A=0.01$ which corresponds to a very small rotational flow parameter, the eigenfunctions still obey a `surface-like' structure, that is, the amplitude of the radial displacement and velocity perturbations possesses a maximum at the boundary of the flux tube and decays away from the boundary. However, increasing the amplitude of the background rotational flow causes the radial displacement perturbation to increase towards the centre of the flux tube, such that the maximum displacement perturbation is no longer at the point where $r=a$. This results in an eigenfunction that shares striking similarities to that of the fundamental kink mode, and may therefore be misinterpreted in observational data.

To further emphasise this point, it is possible to plot the eigenfunctions of $\hat{P}_T$ and $\hat{\xi}_r$ for eigenvalues of a similar phase speed on either side of the modified flow resonance point. One of these solutions corresponds to the slow surface kink mode and the other is the fast surface kink mode. 
\begin{figure*}
    \centering
    \begin{subfigure}{.49\textwidth}
        \centering
        \includegraphics[width=8.7cm]{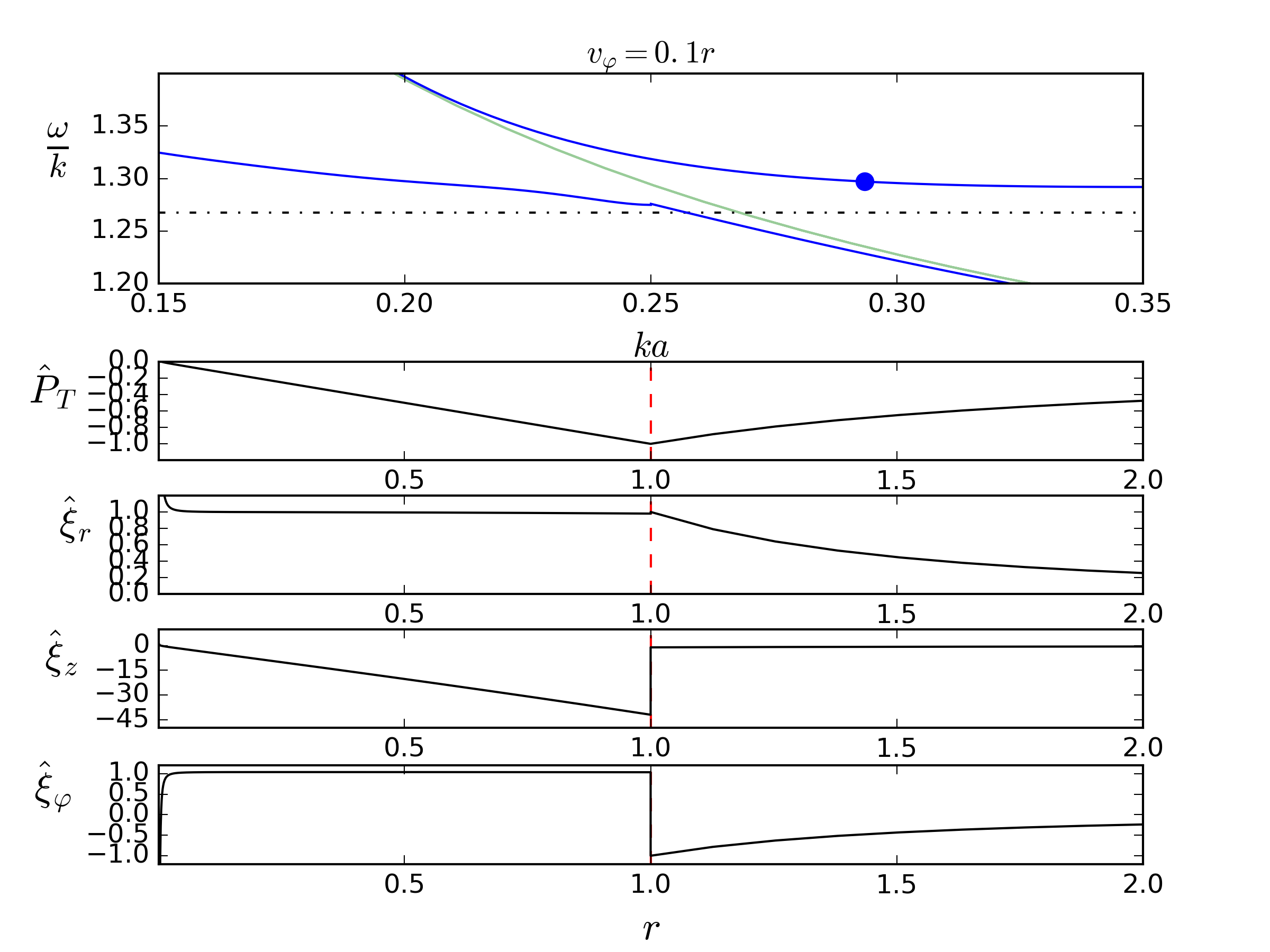}
        \caption{}
        \label{linear_vrot_eigenfunction_comparison_a}
   \end{subfigure}
   \begin{subfigure}{.49\textwidth}
        \centering
        \includegraphics[width=8.7cm]{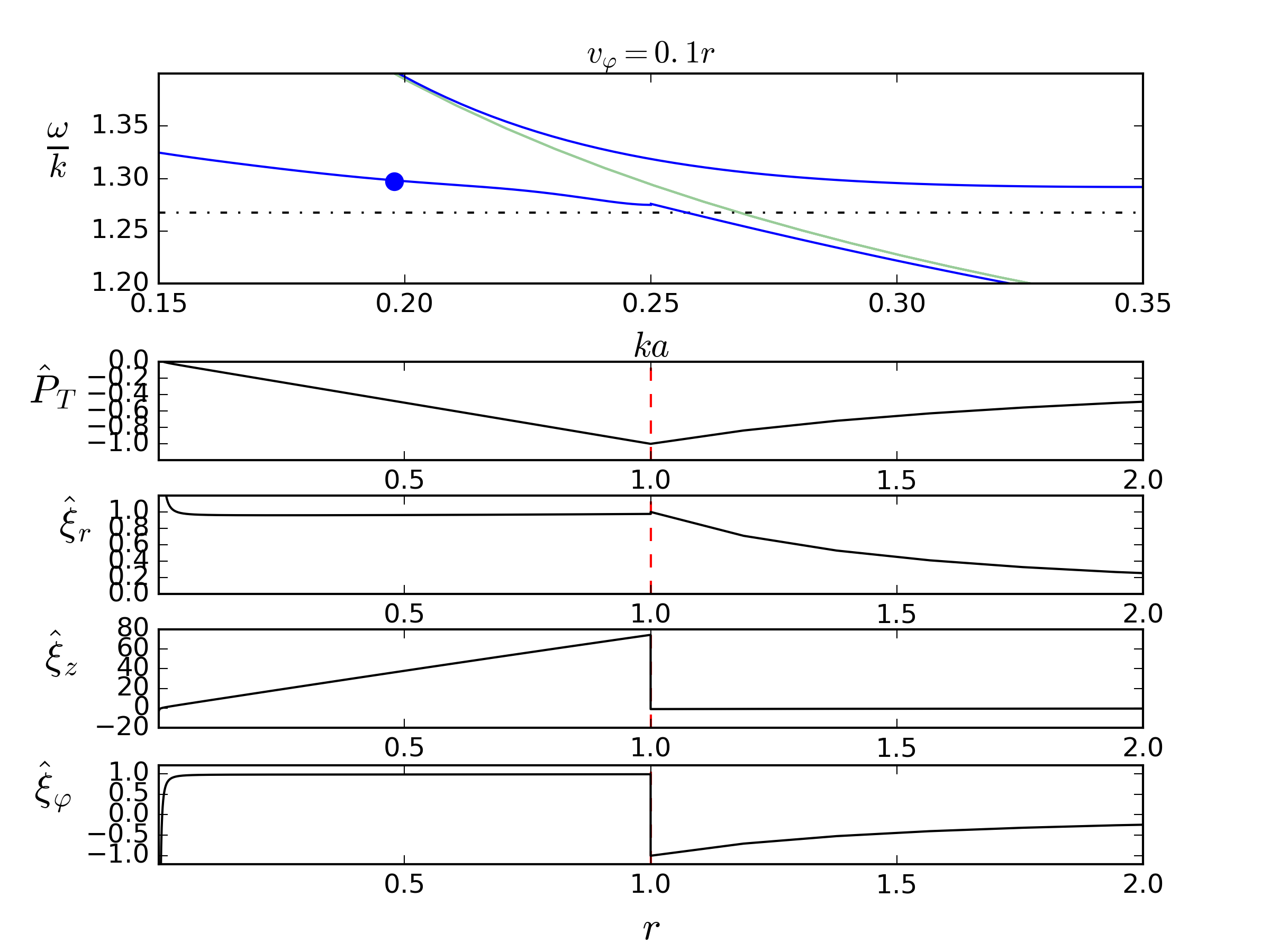}
        \caption{}
        \label{linear_vrot_eigenfunction_comparison_b}
    \end{subfigure} 
    \caption{Panels showing the obtained wave solutions in a zoomed region denoted by the dashed box in Figure \ref{linear_flow_v01_dd}. The eigenfunctions $\hat{P}_T$, $\hat{\xi}_r$, $\hat{\xi}_z$ and $\hat{\xi}_{\varphi}$ for the (a) fast and (b) slow magnetoacoustic surface kink mode solutions in a photospheric flux tube with a background rotational flow given by $v_{\varphi}=0.1r$. Both panels (a) and (b) are for the same $v_{ph}= \omega/k = 1.3$ indicated by the blue dot on the dispersion diagram. All plots are normalised such that the external value of each eigenfunction is equal to unity at the boundary. }
    \label{linear_vrot_eigenfunction_comparison}
\end{figure*}
Shown in Figure \ref{linear_vrot_eigenfunction_comparison} are these eigenfunctions for the slow and fast magnetoacoustic kink modes at a similar phase speed. It can be seen that the normalised eigenfunctions for $\hat{P}_T$, $\hat{\xi}_r$ and $\hat{\xi}_{\varphi}$ are difficult to distinguish between the slow and the fast surface modes. Furthermore, the modes no longer display the typical characteristics of the surface mode anymore. In particular, the main characteristic of a surface mode from the uniform cylinder model is that it possesses a maximum amplitude of radial displacement perturbation at the boundary of the waveguide, which is no longer the case when a rotational background plasma flow is present. Investigating the nature of perturbations parallel and perpendicular to the magnetic field may aid in distinguishing between the slow and fast surface kink modes. Slow modes tend to propagate (mainly) along the magnetic field lines, which in our study are straight and vertical, so we should expect $\hat{\xi}_z$ to dominate for the slow mode when compared to the fast mode, which may propagate at an angle across the magnetic field. In Figure \ref{linear_vrot_eigenfunction_comparison}, we also show the normalised $z$ component of the displacement perturbation for both the fast and slow surface kink modes. As expected, $\hat{\xi}_z$ is dominant over $\hat{\xi}_r$ for the slow surface mode \citep[see e.g.][]{moreels2013}, however, the same is also true for the fast surface mode. Although, the absolute magnitude of the normalised $z$ component is greater for the slow surface mode compared to the fast surface modes, by roughly a factor of $2$, suggesting that the component of displacement is still more dominant for the slow surface kink mode in a rotating photospheric flux tube. It should be noted that, displayed in some plots of the eigenfunctions, the amplitude of the resulting eigenfunctions increases as we approach the center of the cylinder. This is a feature which relates to the fact that there is a singularity in the set of Equations (\ref{rxi_r_diff_eqn}) and (\ref{P_diff_eqn}) at $r=0$. However, this is purely an artifact of the plotting technique and does not affect the eigensolver obtaining the solutions.

\subsubsection{2D and 3D velocity fields}
Following the analysis from Section \ref{Method}, the wave solutions are put proportional to $\text{exp}\left[i\left(m\varphi + kz - \omega t\right)\right]$, therefore, it is possible to convert the one dimensional radial eigenfunctions, for example those shown in Figures \ref{Twisted_flow_kink_eigenfunctions_photospheric} and \ref{linear_vrot_eigenfunction_comparison}, into two and three dimensional plots, to better represent an observer's perspective. 
\begin{figure*}
    \centering
    \includegraphics[width=15.5cm]{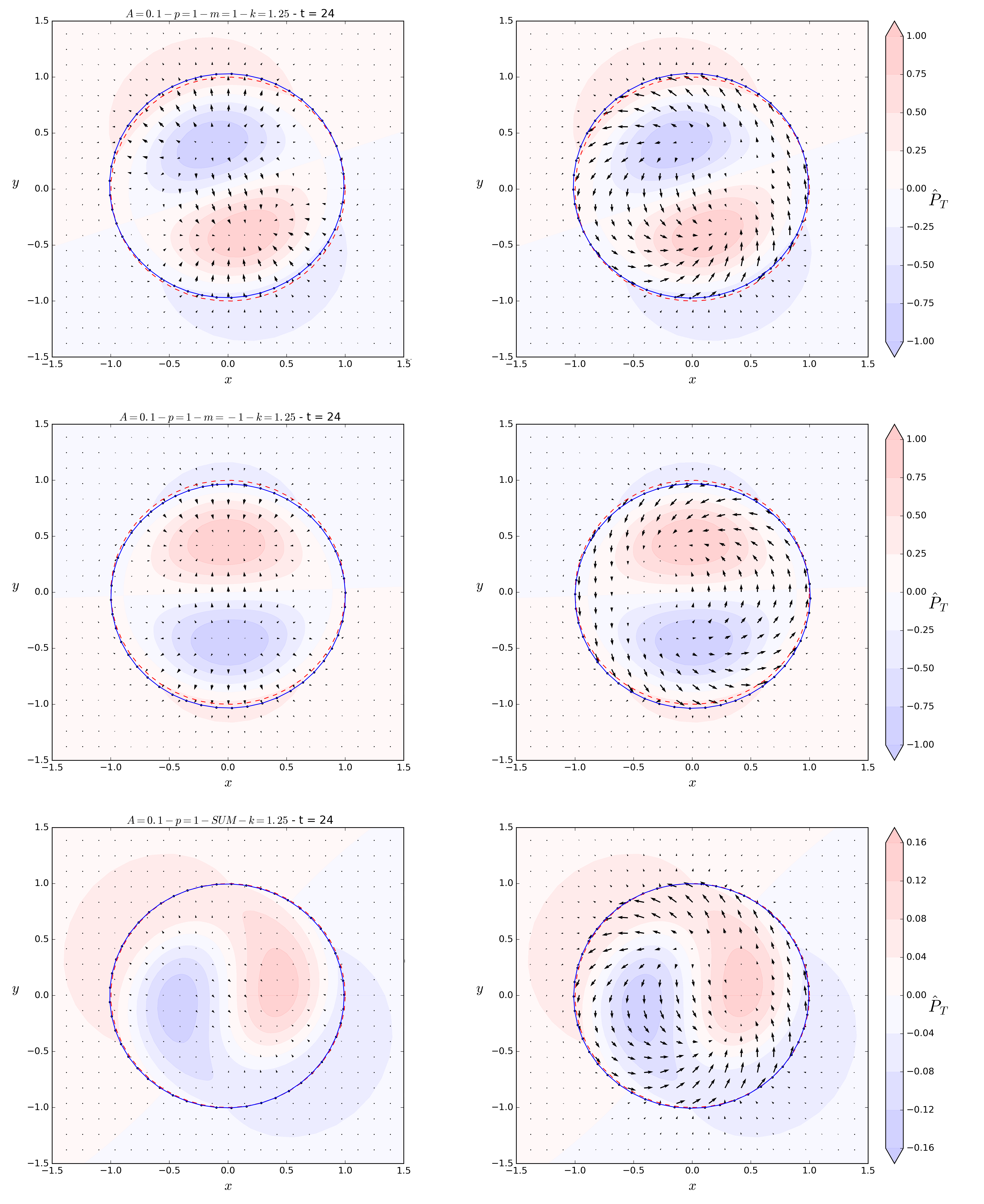}
    \caption{Snapshots of the $2$D velocity field for the scenario of a slow body mode in a photospheric flux in the presence of a background flow given by $v_{\varphi}=0.1r$. The plots are arranged in a $3 \times 2$ configuration where the left hand column shows the velocity perturbation only, whereas the right hand column shows the perturbed velocity field plus the background velocity field. The top row corresponds to the solution for the $m=1$ mode, the middle row shows the solution for the $m=-1$ mode and the bottom row shows the resulting velocity field and total pressure perturbation for the sum of the $m=1$ and $m=-1$ modes. In all panels the velocity vectors are normalised by their maximum values. The colour contour denotes the normalised total pressure perturbation, $\hat{P}_T$, which is the same for both the left and right columns. The boundary of the flux tube is highlighted by the solid blue line. An animated movie of this Figure is available online.}
    \label{2D_moment0}
\end{figure*}
In Figure \ref{2D_moment0}, we show a snapshot of the two dimensional (converted into Cartesian $x$ and $y$) velocity field for the perturbations alongside the addition of the background flow onto the perturbations. This snapshot is chosen specifically at the time when the $m=1$ and the $m=-1$ perturbations effectively cancel each other such that the sum of the perturbed velocity field is zero. Of course, due to the breaking of symmetry from the presence of the background rotational flow, the boundary of the flux tube is no longer at equilibrium, as the $m=1$ and $m=-1$ modes are slightly out of phase with one another. In this snapshot, the perturbed total pressure, displayed by the colour contour, can be seen to display a swirling characteristic, albeit with a small magnitude, and is clearly visible in the bottom panels of Figure \ref{2D_moment0}. This swirling behaviour is present purely in the perturbations, signifying the effect that a rotating waveguide has on the perturbed eigenfunctions. This swirling behaviour of the total pressure perturbation may represent an observational signature of the slow body kink mode propagating in a rotating photospheric flux tube, although detecting this may be a challenge for observers due to the small absolute value of magnitude. In the plots in the right hand column of Figure \ref{2D_moment0}, we add the background velocity field to the perturbed velocity field. The background velocity field is given by $v_{\varphi}=0.1r$ and acts counter-clockwise in the $xy$ plane. The addition of the background velocity field completely changes the observed distribution of the velocity field for the slow body kink mode. In the case of a uniform cylinder, the velocity field for the slow body kink mode can be seen emanating from two `islands', which correspond to two anti-nodes in the total pressure eigenfunction. This can be seen in the left hand column showing the perturbations only for the case of $m=1$ and $m=-1$, however, when the background flow is added, the velocity field no longer displays these typical characteristics. It should be noted that both the perturbed velocity field and the background velocity field are normalised separately to their respective maximum values to aid these visualisations, as it is to be expected that the perturbed components will be significantly smaller in absolute value than the background quantities.

\begin{figure*}
    \centering
    \includegraphics[width=15.5cm]{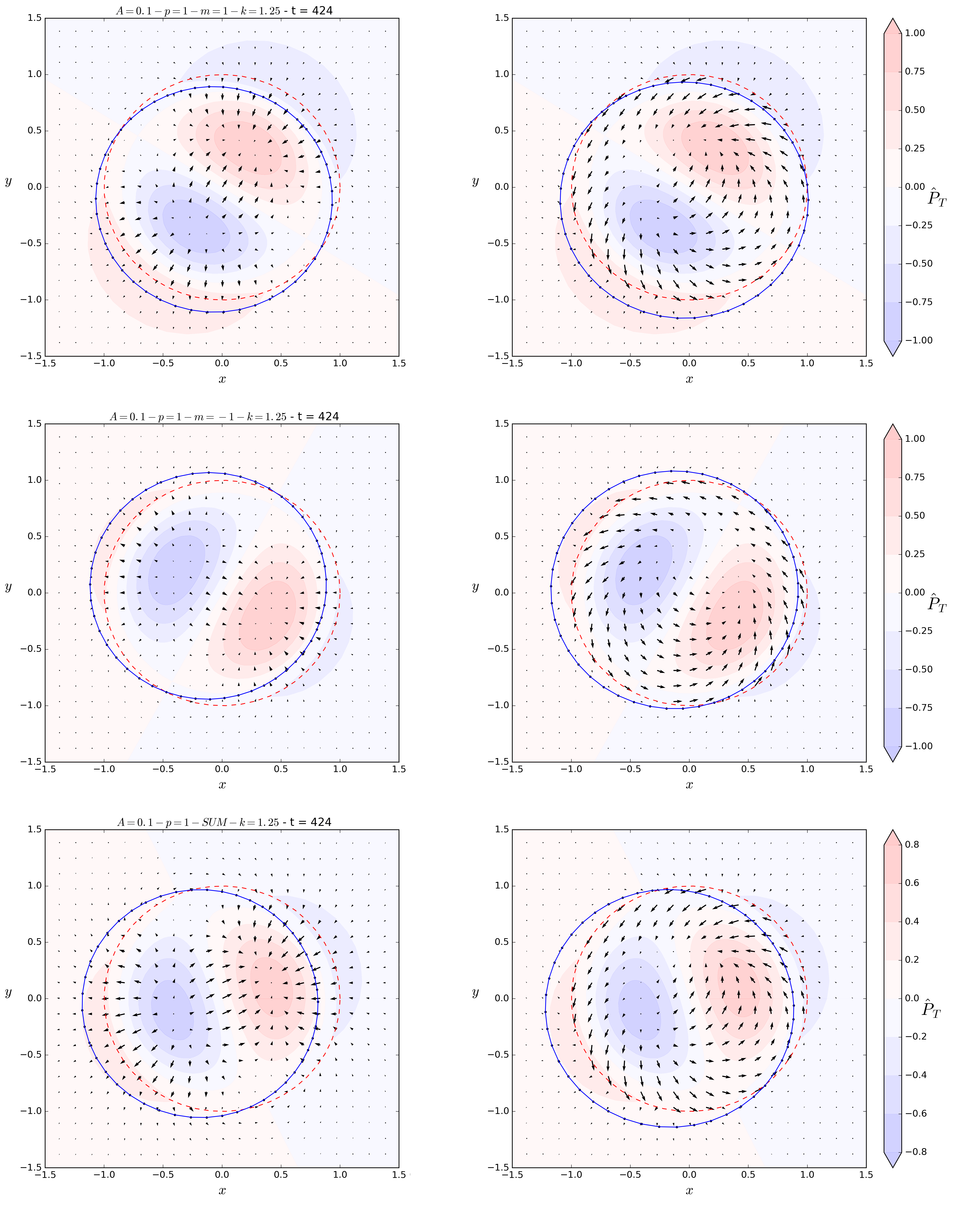}
    \caption{Same as Figure \ref{2D_moment0} but a snapshot at a later time. An animated movie of this Figure is available online.}
    \label{2D_moment1}
\end{figure*}
In Figure \ref{2D_moment1}, we show the same quantities as those displayed in Figure \ref{2D_moment0} at a later time. As we have shown in Figure \ref{m_comparison}, there is a significant difference in the phase speed for the $m=1$ and $m=-1$ forward propagating slow body waves. Therefore, we would expect these perturbations to become increasingly out of phase with one another as time progresses, a behaviour which is displayed in Figure \ref{2D_moment1}. The two modes becoming out of phase with one another can be seen by comparing the top row and the middle rows of Figure \ref{2D_moment1} (which display the $m=1$ and $m=-1$ modes respectively). For the case of a uniform cylinder with no background rotational flow, we would expect that the $m=1$ and $m=-1$ modes would rotate in opposite directions but perfectly in phase with one another, however, this is no longer true when a rotational background flow is added. As a result of one mode rotating with the background flow (in our case this is the $m=1$ mode), and the other mode propagating against the background flow ($m=-1$ mode), the sum of the two modes results in a total pressure perturbation which is also propagating (rotating) in the azimuthal direction. Furthermore, the boundary of the flux tube can be seen to rotate (see associated online movie). The observed rotating motion when combining the individual $m=1$ and $m=-1$ wave modes presents another observational signature that can be sought to identify the kink mode in the presence of a background rotational flow.

   \begin{figure*}
    \centering
    \begin{subfigure}{.99\textwidth}
        \centering
        \includegraphics[width=14.5cm]{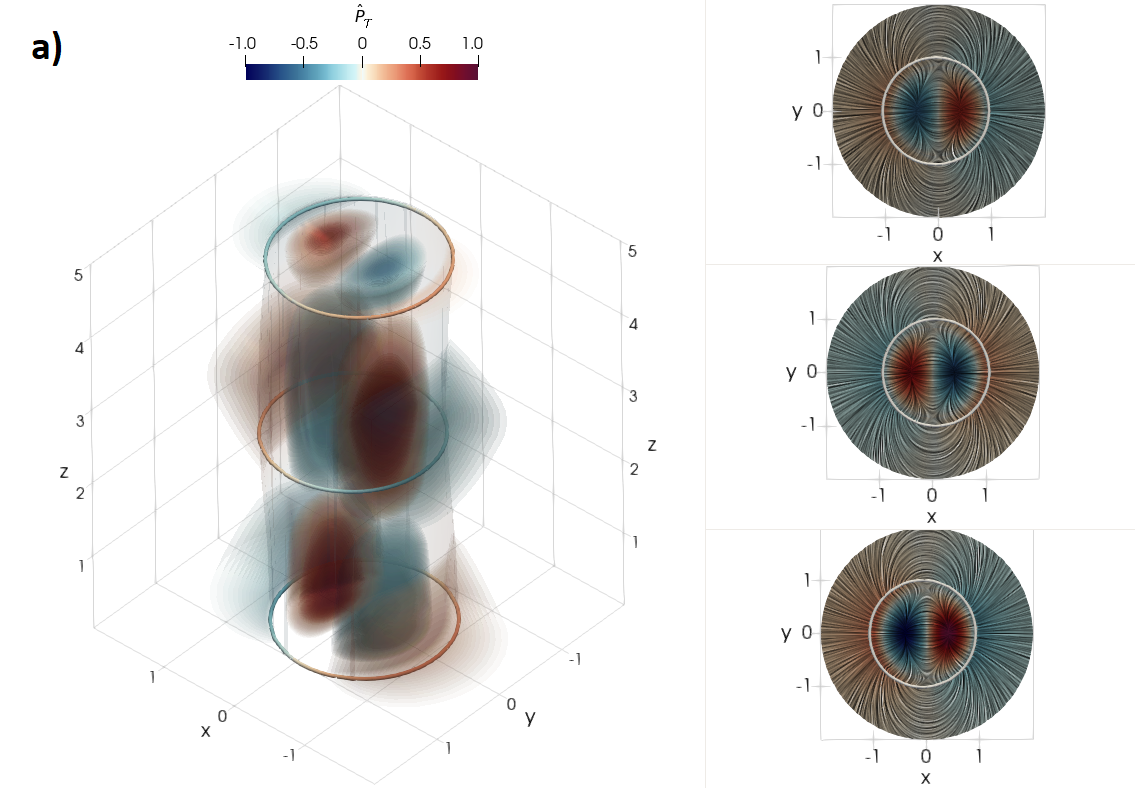}
        \caption{}
        \label{noflow_kink_LIC}
   \end{subfigure}
   \begin{subfigure}{.99\textwidth}
        \centering
        \includegraphics[width=14.5cm]{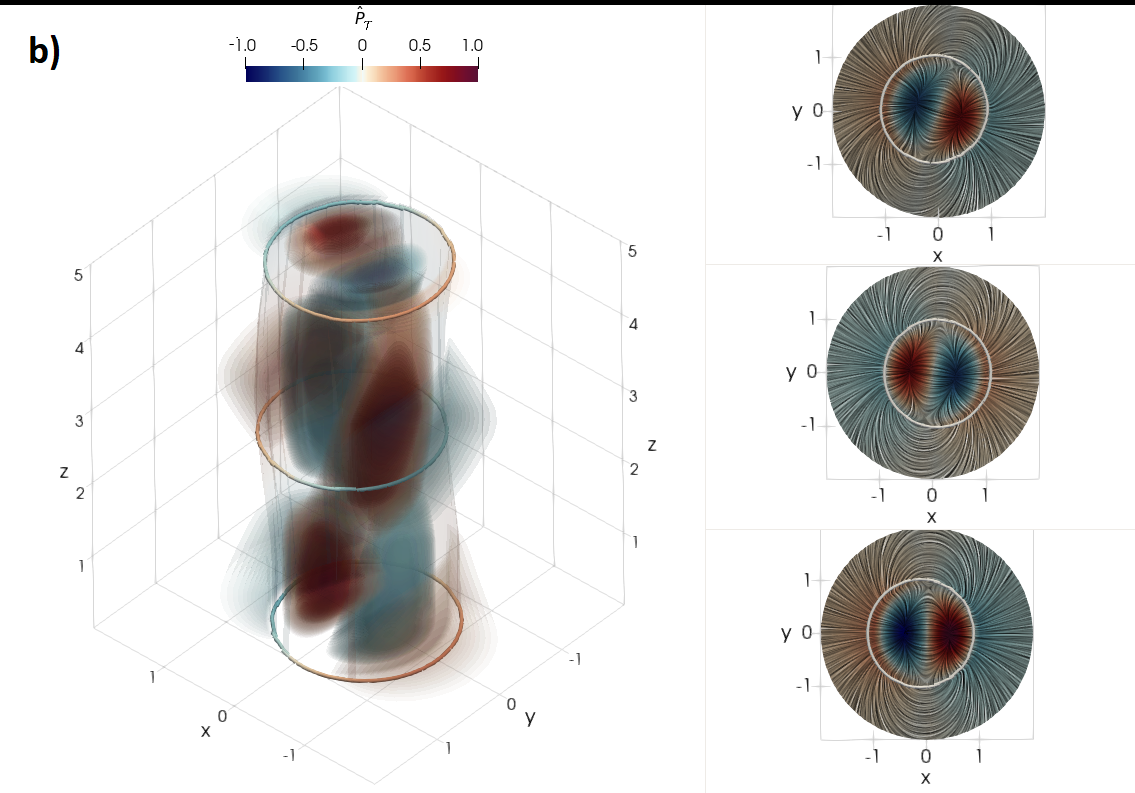}
        \caption{}
        \label{rotflow_kink_LIC}
    \end{subfigure}
    \caption{This figure shows the three dimensional structure of the slow body kink mode for (a) a uniform magnetic flux tube without the presence of a background flow and (b) a magnetic flux tube in the presence of a background rotational flow, by visualising the normalised total pressure perturbation $\hat{P}_T$. On the left panels, the magnetic flux tube is immersed in the volume rendering of $\hat{P}_T$. The three cross-sectional cuts (shown as coloured rings at three different heights, $z=0.0, 2.5$ and $5.0$) correspond to the corresponding right subplots. These three subplots show the LIC visualisation at the same heights. The white rings represent the boundary of the flux tube. An animated version of this Figure is available online.}
    \label{linear_flow_kink_LIC}
   \end{figure*}
Figure \ref{linear_flow_kink_LIC} shows a three dimensional visualisation of how the sum of the $m=\pm 1$ slow body kink mode manifests itself in a uniform magnetic flux tube and also for the case of a magnetic flux tube in the presence of a background rotational flow. Figure \ref{linear_flow_kink_LIC} also displays a Line Integral Contour (LIC) plot at three different heights in the flux tube which are denoted by white rings in the 3D plot. A LIC visualisation aids imaging the velocity field of the plasma \citep[see, e.g.][]{Cab1993}. In other words, LIC visualisations display the path lines that an object would follow if it were placed into the fluid with a stationary velocity field. It can be seen that the LIC for the slow body kink mode in a uniform magnetic flux tube (Figure \ref{noflow_kink_LIC}) remains unchanged with height, as the wave propagates and the structure oscillates in one plane, in other words it is linearly polarised. In addition, the LIC also displays a `double X-point' close to $x=0$ and $y=\pm 1$, near the boundary in the plane in which the structure does not oscillate. This behaviour is due to the rotational motions that are inherently associated with the non-axisymmetric kink mode \citep{goo2014}. However, once a rotational flow is present, the observed evolution of the flux tube motion is affected. Firstly, we can observe that the three dimensional behaviour of the total pressure (density) perturbation begins to represent a helical structure, in contrast to the sinusoidal oscillation in the uniform case, as the perturbation now also displays a rotational behaviour. This suggests that the resulting kink wave is circularly polarised, as a result of the $\pm 1$ modes becoming out of phase with one another. Therefore, we show that, in the presence of a background rotational flow, the kink mode may manifest itself as a circularly polarised mode, similar to a twisted magnetic flux tube \citep{ter2012,rud2015}. 

Furthermore, the LIC visualisations show that the velocity field evolution is modified in the presence of a background rotational flow. Firstly, the `double X-point' feature seen in the LIC for the uniform flux tube no longer displays the same characteristics in the scenario including a background rotational flow. Instead, the `X-points' can be seen at different locations at certain times as the wave propagates upwards. Secondly, the evolution of the velocity field is no longer constant and does not appear to oscillate in a plane on a single axis anymore. Similar to the behaviour seen in Figures \ref{2D_moment0} and \ref{2D_moment1}, the LIC visualisation shows that the velocity field also rotates as it evolves, and the traditional signatures of the kink mode in a uniform flux tube are no longer present. These features could possibly be retrieved from numerical simulations of kink waves in rotating structures.

\section{Conclusions}\label{conclusions}

In the present study, we have extended previous studies investigating the properties of MHD waves in photospheric waveguides, by introducing a linear azimuthal component to the background velocity field. In order to conduct this investigation, we utilised a previously developed numerical eigensolver \citep{Skirvin2021, Skirvin2022} to model a rotating flux tube in a photospheric environment with a background $v_{\varphi}$ component. For the inclusion of a linear rotational flow, very similar results to those of the linear magnetic twist were recovered \citep{ErdFed2010}. We find that the obtained kink mode solutions possess phase speeds along the magnetic field which tend to infinite values in the long wavelength limit where they may become leaky. We have shown this to be a result of the modes being guided by the trajectory of the flow resonance location modified by the slow frequency. As a result of considering a linear background flow, the equations describing the continua regions reduce to single point locations. Therefore, there becomes a point where, under photospheric conditions, the slow surface kink mode and the slow body kink mode approach the same solution at the resonant point and undergo an avoided crossing where their properties become mixed. Comparison of the eigenfunctions for the slow surface kink mode and fast surface kink mode, in the presence of a linear background rotational flow at similar phase speeds, indicated that identification of the two modes becomes extremely difficult in the long wavelength limit using the total pressure and radial displacement perturbations. However, comparison of $\hat{\xi}_z$ shows that this component is still dominant for slow modes, comparable to uniform theory. Furthermore, we find that the axisymmetric $m=0$ sausage mode remains unaffected by any background azimuthal component. Analytically this can be understood by examining the governing set of Equations (\ref{rxi_r_diff})-(\ref{T}) when setting $m=0$ and noticing many terms either simplifying or disappearing altogether. In addition, it suggests that sausage mode observations in the lower solar atmosphere in e.g. pores and sunspots, may not be a suitable wave mode to conduct atmospheric-seismology, if the structure is in the presence of any magnetic twist or background rotational flows.

We have also presented 2D plots showing the velocity field with and without the inclusion of the background flow (Figures \ref{2D_moment0}-\ref{2D_moment1}) and 3D plots showing the perturbation of the normalised total pressure $\hat{P}_T$ (Figure \ref{linear_flow_kink_LIC}). These plots clearly show the breaking of symmetry between the $m=1$ and $m=-1$ kink modes when a flux tube is in the presence of a rotational background flow. Furthermore, inspection of the total pressure perturbations shows that it displays signatures of a swirling motion when the flux tube experiences its maximum displacement. It is hoped that these visualisations will aid future interpretations of observational data from high resolution instruments on state of the art telescopes such as DKIST.
Finally, we have also produced three dimensional visualisations of the kink mode propagating in a photospheric flux tube with a linear background rotational flow incorporated. We have also shown, in Figure \ref{linear_flow_kink_LIC}, the LIC visualisations at different heights displaying the evolution of the velocity field as the wave propagates along the magnetic field, which may be retrieved from numerical simulations.

In this study, we have suggested potential observational signatures of magnetoacoustic kink modes in photospheric flux tubes with background rotational flows. These signatures share striking characteristics with previous observations of circularly polarised kink modes in the lower solar atmosphere \citep{Jess2017, stang2017}. However, naturally, the magnitude of the background flows are much greater than those of the perturbations arising from the waves, which then raises the important question of how to actually observe this phenomena in the solar atmosphere. Both the rotational motion of the waveguide combined with the swirling pattern seen in the total pressure perturbation are observational signatures that may be detected with current and future telescopes, however distinguishing between the contribution of the background and perturbed components may be challenging.

Furthermore, we have focused only on a magnetic flux tube in a photospheric environment. Under the photospheric conditions adopted in this work, the Alfv\'{e}n continuum given by Equation (\ref{alfven_flow_continuum}) exists in the leaky regime at phase speeds above the cut-off speed at $v_{ph} = c_e$. Therefore this resonant region is not discussed in detail in this work, however, it will become important for studies under coronal conditions, where contributions from both Equations (\ref{alfven_flow_continuum}) and (\ref{cusp_flow_continuum}) become important. In the solar atmosphere, vortices can be buffeted from multiple directions, for example convective motion due to convection cells. In this study, we have considered just one fixed driver that oscillates from side to side, however in reality the picture will be more complicated due to drivers in additional directions. These perturbations along different axis will manifest themselves as apparent rotation, regardless of a background rotational flow, this is work to be investigated in the future. Additional future work includes a closer inspection on the effect that rotational flows have on the sausage mode. We have shown that the sausage mode may be affected by the presence of background rotational flow in magnetic flux tubes, however, more insight is required to quantify exactly how the sausage mode is affected and how it manifests itself in rotating flux tubes. Finally, future work to be investigated includes modelling a background rotational flow with a realistic radial profile. For example \citet{Silva2020} have shown, using information from MURAM simulations, that the radial profile of the azimuthal velocity component in solar vortex tubes can be accurately modelled with a cubic polynomial. In this case, the background rotational flow becomes non-linear, such that the value of parameter $\alpha$ in Equation (\ref{flow_eqn}) no longer equals $1$. We expect that in this example, the MHD spectrum is densely occupied by the (modified) flow continua, consequently, trapped modes may find it difficult to exist. Obtaining the eigenvalues that lie inside the continua can be achieved by modifying the numerical eigensolver employed in this work or by using an alternative 1D eigensolver such as Legolas \citep{claes2020}. Modelling a non-linear background rotational flow such as this is an objective for future studies.

\section*{Acknowledgements}
This work has been supported by STFC (UK). SJS is grateful to STFC for the PhD studentship project reference (2135820). VF, GV, and SSAS are grateful to Science and Technology Facilities Council (STFC) grant ST/V000977/1, and The Royal Society, International Exchanges Scheme, collaboration with Brazil (IES191114) and Chile (IE170301). TVD was supported by the European Research Council (ERC) under the European Union's Horizon 2020 research and innovation programme (grant agreement No 724326) and the C1 grant TRACEspace of Internal Funds KU Leuven. The results received support from the FWO senior research project with number G088021N. SJS and VF would like to thank the International Space Science Institute (ISSI) in Bern, Switzerland, for the hospitality provided to the members
of the team on ‘The Nature and Physics of Vortex Flows in Solar Plasmas’. SJS and GV wish to acknowledge scientific discussions with the Waves in the Lower Solar Atmosphere (WaLSA; https://www.WaLSA.team) team, which is supported by the Research Council of Norway (project no. 262622). This research has also received financial support from the ISEE, International Joint Research Program  (Nagoya University, Japan) and the European Union’s Horizon 2020 research and innovation program under grant agreement No. 824135 (SOLARNET).

\textit{Software}: Numpy \citep{harris2020}, matplotlib \citep{Hunter2007}, SciPy \citep{Vir2020}. The numerical code, Sheffield Dispersion Diagram Code (SDDC) introduced and applied in this work is available on the Plasma Dynamics Group (PDG) website \footnote{\url{https://sites.google.com/sheffield.ac.uk/pdg/solar-codes?authuser=0}} along with the user manual which explains some cases shown in this work. This code and the accompanying tools have been developed using Python an open-source and community-developed programming language.

\section*{Data Availability}
Accessibility of data used in this research is available upon request from the authors.



\bibliographystyle{mnras}
\bibliography{ref} 







\bsp	
\label{lastpage}
\end{document}